\documentclass[twocolumn,amsmath,amssymb,aps,longbibliography,floatfix]{revtex4-1}

\usepackage{cancel}
\usepackage{enumitem}
\usepackage[utf8x]{inputenc}
\usepackage{graphicx}
\usepackage{dcolumn}
\usepackage{bm}
\usepackage{hyperref}
\usepackage[switch]{lineno}
\usepackage{float}             
\usepackage{subfigure}
\usepackage{tikz}

\begin{document}

\title{Surfing multiple conformation-property landscapes \textit{via} machine learning: Designing magnetic anisotropy.}

\author{$^{1}$Alessandro Lunghi}
\email{lunghia@tcd.ie}
\author{$^{1}$Stefano Sanvito}

\affiliation{$^{1}$School of Physics, AMBER and CRANN, Trinity College, Dublin 2, Ireland}

\begin{abstract}
{\bf The advent of computational statistical disciplines, such as machine learning, is leading to a paradigm shift 
in the way we conceive the design of new compounds. Today computational science does not only provide a sound 
understanding of experiments, but also can directly design the best compound for specific applications. This approach, 
known as reverse engineering, requires the construction of models able to efficiently predict continuous structure-property 
maps. Here we show that reverse engineering can be used to tune the magnetic properties of a single-ion molecular 
magnet in an automated intelligent fashion. We design a machine learning model to predict both the energy and 
magnetic properties as function of the chemical structure. Then, a particle-swarm optimization algorithm is used
to explore the conformational landscapes in the search for new molecular structures leading to an enhanced magnetic 
anisotropy. We find that a 5\% change in one of the coordination angles leads to a $\sim$50\% increase in the anisotropy. 
Our approach paves the way for a machine-learning-driven exploration of the chemical space of general classes of 
magnetic materials. Most importantly, it can be applied to any structure-property relation and offers an effective way 
to automatically generate new materials with target properties starting from the knowledge of previously synthesized 
ones.}
\end{abstract}

\maketitle

The design of new materials with specific target properties is the ultimate goal of reverse materials engineering. 
This approach requires the construction of a range of maps between the chemical structure $\{r_{i},Z_{i}\}$ and 
the properties of interest $P(\{r_{i},Z_{i}\})$\cite{Franceschetti1999}. In this framework, the design process corresponds to the optimization 
of a global target function, $\chi$, that weighs different properties,
\begin{equation}
\max_{\{r_{i},Z_{i}\}} \chi=\sum_{l}\gamma_{l}P(\{r_{i},Z_{i}\})\:,
\label{genmap}
\end{equation}
where $\{r_{i},Z_{i}\}$ contains the position and atomic number of the atoms forming the material and the weights, 
$\gamma_{l}$, set the relative importance of the single properties, $P_{l}$.

While in principle quantum mechanical methods could be used for such a task, their computational overheads render 
them impractical. Machine-learning-based models, with their ability to reproduce quantum mechanical results at a
negligible computational cost, are the perfect tool to construct reverse engineering and generative 
approaches~\cite{Ramprasad2017,Gomez-Bombarelli2018,Sanchez-lengeling2018,Butler2018}. In this work we develop 
an efficient framework to build and explore general properties-structure maps and apply it to a topical case of technological 
importance, namely magnetic materials.

Magnetism is an exotic phenomenon that emerges from a very delicate balance between the electronic and structural 
properties of chemical compounds. Magnetic compounds form a paradigmatic materials class containing rare members
with large technological impact. The working principle of hard magnets is based on the presence of a large axial magnetic 
anisotropy that stabilises the magnetic moment against thermal fluctuations. This picture can be formally explained with 
the spin Hamiltonian
\begin{equation}
\hat{H}_{S}=D\hat{S}^{2}_{z}\:,
\end{equation}
where $\hat{S}_{z}$ is the $z$ component of the spin operator and a negative $D$ parameter corresponds to an axial 
anisotropy. When the magnetic anisotropy is not large enough to overcome thermal fluctuations or when a magnet is 
perturbed by external stimuli, demagnetization processes take place. Spin relaxation in non-metallic materials is ultimately
due to the interaction between the magnetic and lattice degrees of freedom, namely the spin-phonon coupling. This 
interaction manifests itself through the dependence of the magnetic anisotropy tensor $\mathbf{D}(r_{i})$ on the 
atomic positions $r_{i}$. In general $\mathbf{D}(r_{i})$ is interpreted in terms of a Taylor expansion around the equilibrium 
molecular geometry $r_{i,0}$:
\begin{equation}
\mathbf{D}(r_{i}-r_{i,0})=\mathbf{D}+\sum_{i}\left(\frac{\partial \mathbf{D}}{\partial r_{i}}\right)_{0}(r_{i}-r_{i,0})+...
\label{taylor}
\end{equation}
The understanding of the microscopic processes leading to spin-phonon coupling and demagnetization is of interest for 
several applications such as heat-assisted magnetic recording\cite{Kryder2008}, ultra-fast demagnetization\cite{Mueller2011}, 
magnetostriction\cite{Slonczewski1961}, spintronics\cite{Maehrlein2018}, molecular magnetism\cite{Lunghi2017,Escalera-moreno2018} and quantum computing based on electronic spins\cite{Whiteley2019,Lunghi2019b}.

The design of magnetic materials is a prototypical example where multiple features must be optimized at the same time 
to reach optimal efficiency. Eq.~(\ref{taylor}) suggests that making high-temperature hard-magnets requires: i) the maximization 
of the axial magnetic anisotropy $|D|$, ii) the minimization of its derivatives (the spin-phonon coupling coefficients), and 
iii) little thermally populated molecular motions $(r_{i}-r_{i,0})$. This effectively corresponds to the design of $\mathbf{D}(r_{i})$ 
having a maxima in correspondence of the equilibrium geometry and a very stiff material~\cite{Lunghi2017}. 

\begin{figure}
\centering
\includegraphics[scale=1]{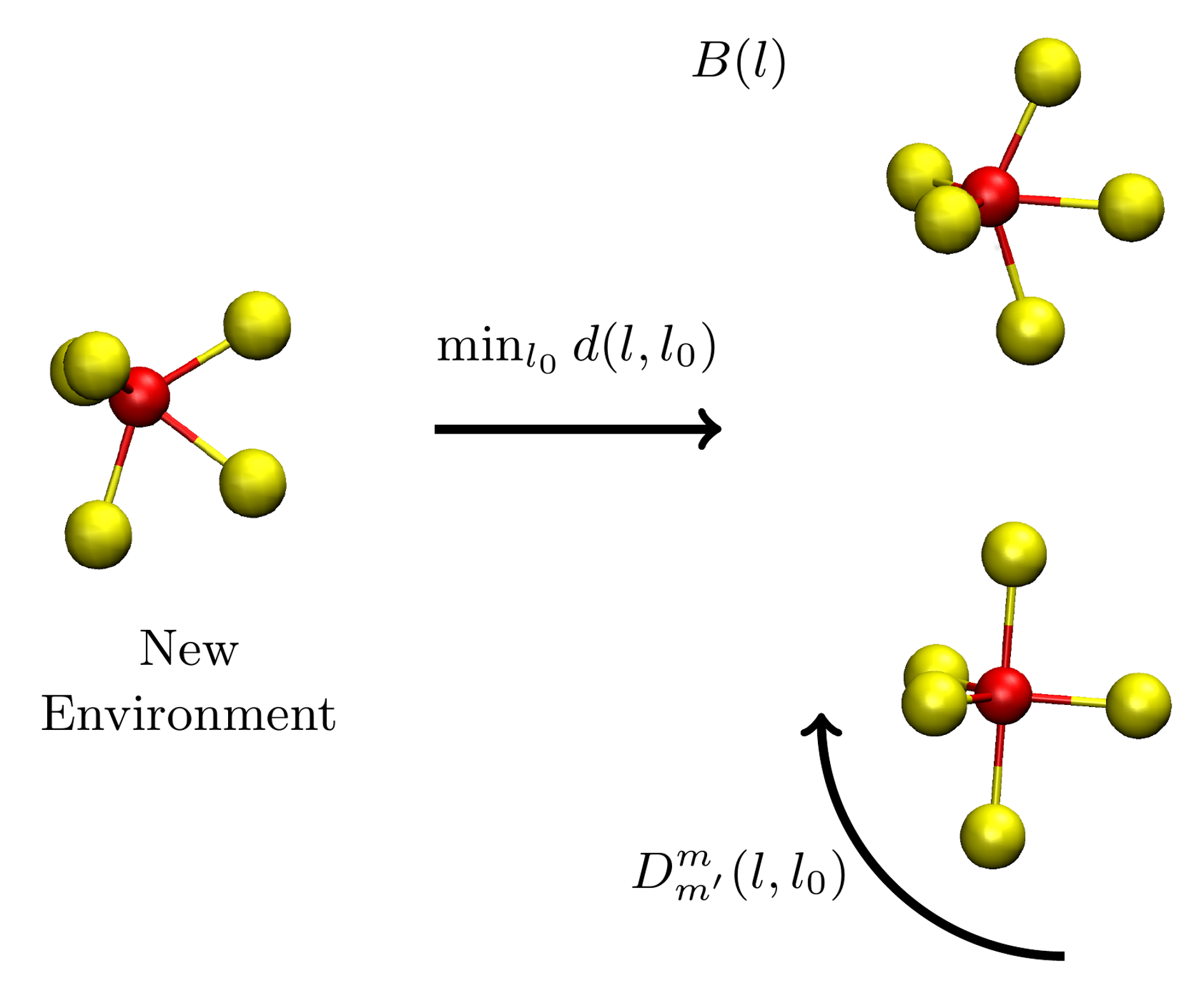}
\caption{\textbf{Scheme for Structural Recognition and Decomposition.} A generic [FeCl$_{5}$]$^{2-}$ distorted structure 
is first compared to all the reference geometries by means of the metric $d(l,l_{0})$. Once the appropriate reference orientation 
is chosen, the structure is decomposed into its internal and rotational contributions. }
\label{scheme}
\end{figure}

We have applied our method to molecular magnets, which appear as the ideal materials class for a systematic design strategy. 
These compounds represent the ultimately small building blocks of magnetic recording media~\cite{Sessoli1993}. At the same 
time the combination of density functional theory (DFT) and post Hartree-Fock methods is ideal for predicting their structural 
and magnetic properties~\cite{Neese2019}. Finally, and most importantly, the extensive synthetic versatility of such coordination 
compounds allows one to fine tune the structure, so that design is practically possible.

Our machine learning strategy is based on Ridge regression and bi-spectrum components as molecular geometry fingerprints~\cite{Bartok2010,Thompson2014,Lunghi2019}. The first step requires the decomposition of the magnetic 
anisotropy over atomic contributions. It is then convenient to write $\mathbf{D}(r_{i})$ over a basis of second-order spherical tensors, $T^{m}$, where $m$ is one of the five spherical tensor components needed to describe a trace-less symmetric second-order Cartesian tensor like $\mathbf{D}$. The explicit relation between $T$ and $\mathbf{D}$ is provided as Supplementary Information. For a molecule containing $N_\mathrm{a}$ atoms the decomposition reads
\begin{equation}
T^{m}=\sum_{i}^{N_\mathrm{a}}T^{m}(i)=\sum_{i}^{N_\mathrm{a}}\sum_{j}^{N_{j}}\alpha^{m}_{j}(i)B_{j}(i)\:,
\label{covridge}
\end{equation}
where the index $j$ runs over the $N_{j}$ bi-spectrum components, $B_j$, describing the atomic environment 
of the $i$-th atom, and $\alpha^{m}_{j}$ are the coefficients that need to be determined through Ridge regression.

Magnetic anisotropy is a tensor quantity, so that Eq.~(\ref{covridge}) needs to be recast in a covariant form to 
ensure that the correct rotational symmetries are enforced. Similar concepts have been recently applied in the 
context of Kernel regression for the prediction of atomic forces and general tensor quantities~\cite{Glielmo2017,Grisafi2018}. 
Since the $B$ terms are rotationally invariant, this is achieved by requiring the coefficients $\alpha_{j}$ to transform 
as spherical tensors with respect to a reference frame rotation,
\begin{equation}
\alpha_{j}^{m}(i)=\sum_{m'}W^{m}_{m'}(l,l_{0})\alpha_{j}^{m'}(i_{0})\:,
\label{rotwig}
\end{equation}
where $W^{m}_{m'}(l,l_{0})$ is the Wigner matrix corresponding to a rigid rotation of the atomic environment of the $l$-th atom with respect to the atomic environment state $l_{0}$ of the same atom chosen as reference orientation. The atom $l$, which defines the local environment used to perform the rotation, is chosen on the basis of the property to model. In the case of the magnetic anisotropy it is the magnetic element. It is important to remark that the choice of a local atomic environment as reference orientation, instead of the entire molecular structure, is fundamental in order to maintain the local nature of the properties learned by the model. For instance, in the case of magnetism, this avoids the issue of accounting for a spurious rotation of the tensor $T^{m}$ when only atoms far away from the magnetic atom are moving.
Once the Ridge regression has determined the unknown coefficients $\alpha_{j}^{m'}(i_{0})$, equations (\ref{covridge}) and (\ref{rotwig}) can be used to predict the magnetic anisotropy for a new configuration. Its intra-molecular geometry is described by the bi-spectrum components, $B_{j}$, and its orientation in space by a Wigner matrix, $W^{m}_{m'}$. This approach is completely general for atomic-local quantities and can include several orientation reference states to describe different coordination environments. 

In order to illustrate our approach we apply it to two typical coordination environments: bi-pyramidal [FeCl$_{5}$]$^{2-}$ 
and trigonal prismatic [FeCl$_{6}$]$^{3-}$. For each molecule we prepare 700+700 configurations, where all the 
Cartesian coordinates of all the atoms are displaced by a random 
value within the limits $\pm$0.1 \AA\ and $\pm$0.2 \AA$ $. The size of the maximum displacements is chosen to be large enough to 
guarantee a broad sampling of out-of-equilibrium configurations, while maintaining a sensible chemical structure. These random displacements 
are applied to the DFT-optimized geometry. We use CASSCF(5,5) to compute the magnetic anisotropy for each of these 1,400 configurations. 
A total of 400 prototypes are excluded from the training set and left for validation and testing purpose. Details on the ab initio calculations are 
provided as Supplementary Information.

In general, the covariance of the Ridge regression is imposed by enforcing the correct rotational properties to the regression's coefficients. 
This can only be done when a rotation between structures is well defined. Thus, structures with different number of atoms and different chemical 
species must belong to different references. Since the two molecules contain Fe$^{+3}$ in two distinct atomic environments with 
different numbers of atoms, one reference orientation per molecule is needed. Moreover, since magnetism is a local property of the Fe atoms, 
the Wigner matrices used to impose the covariance condition are calculated with respect to the Fe atomic environment. In particular, the Fe 
atomic environments of the FeCl$_{5}$ and FeCl$_{6}$ optimized geometries are chosen as the reference atomic environments appearing in Eq.~(\ref{rotwig}).

For each new molecular configuration, it is possible to automatically select one of the two reference orientations by introducing a norm function $d(l,l_{0})=\sum_{j}^{N_{j}}|B_{j}(l)-B_{j}(l_{0})|^{2}$ that points the reference atom $l$ to the optimal reference state on the basis of atomic environment similarities, i.e. the optimal reference configuration $l_{0}$ is represented by the one that minimizes the norm $d(l,l_{0})$. In the case of FeCl$_{5}$ and FeCl$_{6}$ this trivially corresponds to selecting the reference atomic environment with the correct number of atoms. However, the procedure is general and can be applied to any number of reference configurations. Once the correspondence between a molecular configuration and a reference environment is established, the rotation between the two can be computed by applying the Eckart-Sayvets conditions to the Cartesian displacements of the two set of coordinates~\cite{Neto2006}.

To summarize, the procedure that lead to the calculation of a local property of atom $l$ through Eq.~(\ref{rotwig}) involves the following steps: i) comparison of the local atomic environment of atom $l$ with the selected reference local atomic environments $l_{0}$, ii) calculation of the amount of rotation between the local atomic enviornment and the reference local atomic environments that minimizes $d(l,l_{0})$, iii) calculation of $W_{m'}^{m}$, iv) calculation of the bi-spectrum components of all the atoms in the local environment of the atom $l$ and v) calculation of Eq.~(\ref{rotwig}). A schematic representation of the orientation selection process, followed by the rotational and intra-molecular structural decomposition, is provided in Fig.~\ref{scheme}.

\begin{figure}[h!]
\centering
\includegraphics[scale=1]{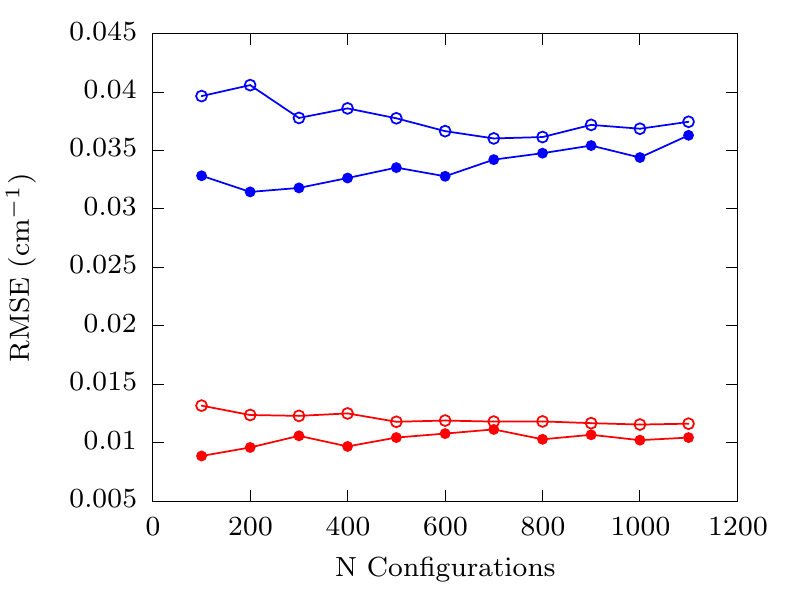}
\caption{\textbf{Magnetic anisotropy training curve for [FeCl$_{5}$]$^{2-}$ and [FeCl$_{6}$]$^{3-}$ complexes.} The RMSE 
between the CASSCF anisotropy and the one predicted by machine learning for the [FeCl$_{5}$]$^{2-}$ and [FeCl$_{6}$]$^{3-}$ 
complexes is plotted as function of the number of configurations included in the training set. The case where covariance is enforced 
is displayed by red curves and symbols, while the case where covariance is not enforced is in blue. Full symbols are used for the 
RMSE of the training set and empty symbols for the test set.}
\label{Fefit}
\end{figure}

In order to illustrate the importance of imposing the covariance property to the Ridge regression we perform the training 
of a model with and without its enforcement. For this purpose the configurations in the training and test sets were rotated 
along the $y$ direction by a random angle in the range $[-45^{\circ}:45^{\circ}]$. Results are reported in Fig. \ref{Fefit} 
and demonstrate the improvement of the covariant method over the non-covariant one. Fig. \ref{Fefit} also demonstrates the high learning rate of the model that has already achieved a converged root mean square error (RMSE) of $\sim 0.01$ cm$^{-1}$ in about 100 configurations/molecule. In comparison, the magnetic anisotropy for [FeCl$_{5}$]$^{-2}$ and [FeCl$_{6}$]$^{-3}$ ranges between 0.05 and 0.3 cm$^{-1}$.

\begin{figure}[h!]
\centering
\includegraphics[scale=1]{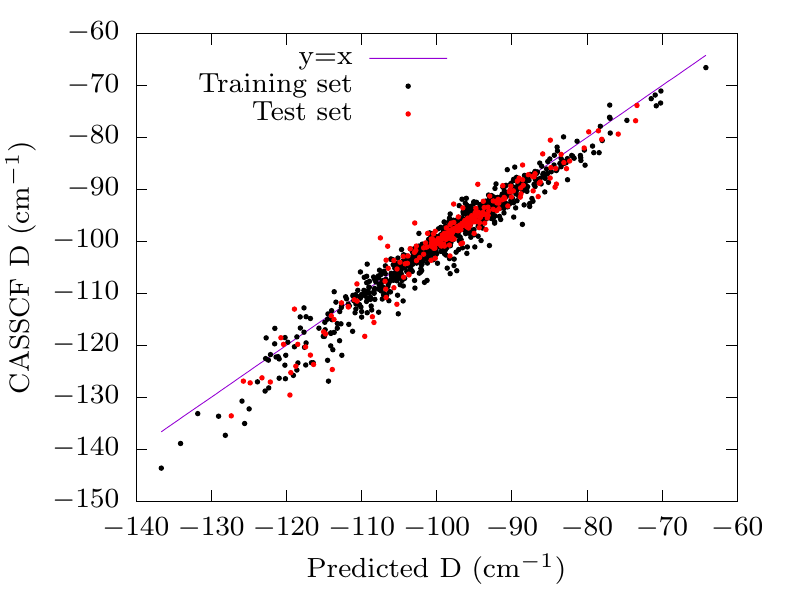}
\includegraphics[scale=1]{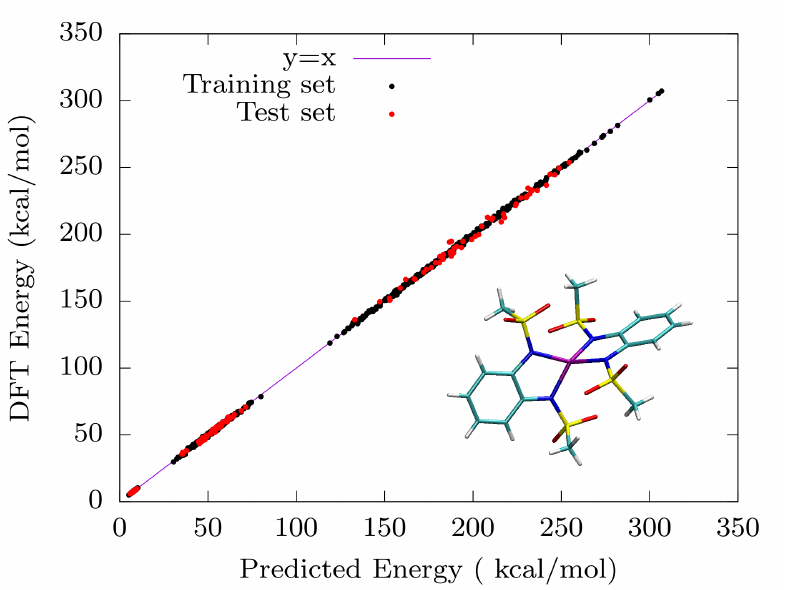}
\caption{\textbf{Magnetic anisotropy and conformational energy training curves for [Co(pdms)$_{2}$]$^{2-}$.} The top panel 
shows the comparison between reference and predicted values of the axial anisotropy $D$, while the bottom panel reports 
the results for the conformational energy. Black dots corresponds to the training set and the red dot corresponds to the test 
one. The inset shows the molecular structure of [Co(pdms)$_{2}$]$^{2-}$, where the Co atom is coloured in Purple, Carbon atoms in Green, Hydrogen atoms in White, Sulphur atoms in Yellow, Nitrogen atoms in Blue and Oxygen atoms in Red.}
\label{Cofit}
\end{figure}

Next we want to demonstrate that our strategy works for real systems and that can be effectively used to explore the magnetic-anisotropy
landscape. To this end we show results for one of the top-performance high-anisotropy single-ion magnet 
[Co(pdms)$_{2}$]$^{2-}$, where pdms=1,2-bis(methanesulfonamido)benzene~\cite{Rechkemmer2016}. As shown in the inset of 
the bottom panel of Fig.~\ref{Cofit}, the Co$^{2+}$ ion is tetrahedrally coordinated by RN$^{-}$ ligands. We optimize the structure in vacuum and 
use it to generate 500 configurations with maximum displacements of $\pm$0.05 \AA$ $, 500 configurations with maximum displacement of $\pm$0.1 \AA$ $ and 500 configurations with maximum displacement of $\pm$0.2 \AA$ $. We then retaine 600 of them for validation and testing purposes. For each configuration we use DFT and CASSCF to compute energy 
and magnetic anisotropy, respectively. More details on the construction of the bispectrum components and the regression are provided as Supplementary Information. Figure~\ref{Cofit} shows the regression results for both axial anisotropy $D$ and conformational energy. The test sets' RMSE measure 1.6 kcal/mol and 2.2 cm$^{-1}$, respectively. Additional 131 configurations have been self-consistently sampled by molecular dynamics in the range 100~K - 400~K to enforce structural stability, as discussed previously\cite{Lunghi2019}. After the inclusion of these configurations the training set's RMSE increases from 1.00 to 3.3 kcal/mol.

The ability to reconstruct continuous structure-energy and structure-magnetic anisotropy maps with virtually no computational 
effort opens now the possibility to select new molecular conformations with optimal properties. We implement a particle-swarm 
optimization (PSO) algorithm\cite{Kennedy1995} and perform a global optimization of the function in Eq.~(\ref{genmap}), specialized to the specific case,
\begin{equation}
\chi(r_{i})=E(r_{i})+\gamma D (r_{i})\:.
\label{pso}
\end{equation}
Here the energy of a molecular conformation $E(r_{i})$ and the magnetic axial anisotropy $D(r_{i})$ are the features of interest. Large 
values of the parameter $\gamma$ would allow for more severe distortions of the equilibrium molecular geometry in favour 
of a more favourable anisotropy. Even though magnetic anisotropy is the relevant figure of merit, the inclusion of energy in 
the target quantity of Eq.~(\ref{pso}) is of fundamental importance as it imposes the exploration of only those low-energy 
molecular distortions that are strongly coupled with $D(r_{i})$. This analysis automatically reveals \textit{all} the magneto-structure correlations relevant to the spin relaxation process, as suggested by Eq.(\ref{taylor}). At the same time, the inclusion of energy in the function $\chi$ also excludes the sampling of totally unrealistic molecular conformations, which would however maximize the anisotropy.

\begin{figure}[h!]
\centering
\includegraphics[scale=1]{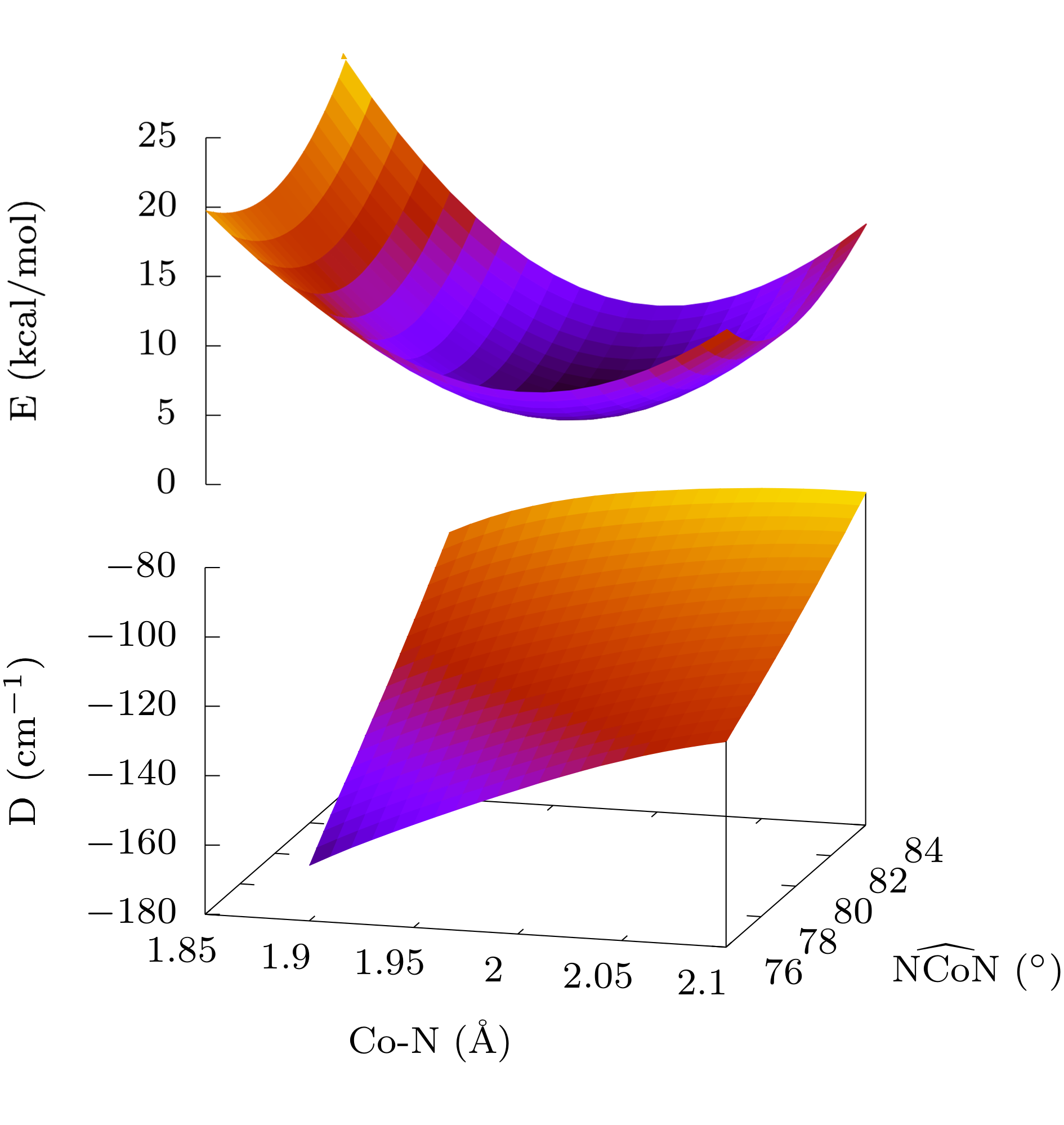}
\caption{\textbf{2D scan of the [Co(pdms)$_{2}$]$^{2-}$ magnetic anisotropy and energy.} The axial magnetic anisotropy 
and the conformational energy, reported in cm$^{-1}$ and kcal/mol respectively, are scanned along the Co-N and 
$\widehat{\textrm{NCoN}}$ directions.}
\label{Coscan}
\end{figure}

The optimization of Eq.(\ref{pso}) shows that the magnetic anisotropy is strongly enhanced by the reduction of Co-N distances and the $\widehat{\textrm{NCoN}}$ angles belonging to the same pdms ligand. Running the PSO for different values of $\gamma$ always lead to the same simple structural distortion. It is important to remark that the the ML model and the PSO exploration are extended to all the molecular degrees of freedom and without any implicit small-displacement regime constraint. This makes it possible to conclude that the simple magneto-structural correlation we found is the only relevant one for this specific chemical environment. The only restriction to the method is imposed by the ability of ML to make accurate predictions for geometries not included in the training set. This issue can be easily contained by implementing an active-learning scheme like the one we used to generate part of the training set with molecular dynamics\cite{Lunghi2019}. In terms of efficiency it is worth remarking that the PSO optimization requires the evaluation of the function $\chi$ at least 1000 times. This means that a comprehensive exploration of the conformational space such as the one presented here is not compatible with a brute-force use of electronic structure methods.

Fig.~\ref{Coscan} shows the 2-dimensional scan of the molecular anisotropy and energy along a 21x21 homogeneous grid of Co-N distances and the $\widehat{\textrm{NCoN}}$ angles. An additional CASSCF calculation of the molecular anisotropy for the geometry corresponding to the largest absolute value of $D$ explored in Fig. \ref{Coscan} confirms a good extrapolating accuracy of the ML outside the original training set with an error of $\sim 7$\%. A large spin-phonon coupling is observed for the individuated molecular motions, where a 5\% reduction of the structural parameters leads to a 50\% increase in the molecular anisotropy. The unexpected simplicity of this magneto-structural correlation is particularly favourable as it suggests a simple rule of thumb to chemically engineer new Co$^{2+}$ single ion magnets with tetrahedral coordination. Our results are in perfect agreement with recent literature that reports the $\widehat{\textrm{NCoN}}$ angles as one of the main handle for improving Co$^{+2}$ single ion magnets~\cite{Fataftah2014,Carl2015,Rechkemmer2016,Suturina2017,Chen2019}. It is fundamental to remark that this result has been achieved without any bias coming from experimental studies and in a complete automatic fashion. In contrast, the experimental derivation of similar magneto-structural correlations generally take significant efforts and a cohort of different experimental characterization techniques~\cite{Rechkemmer2016,Suturina2017}. Moreover, while here we are scanning the entire conformational space, experiments usually explore no more than a couple of degrees of freedom at the time. In this respect, it is not surprising that the Co-N distance magneto-structural correlation has never been reported before as little or no control is applicable to this geometrical parameter at the synthetic level. 

Figure~\ref{Coscan} also shines new light on the nature of the spin-phonon coupling in highly anisotropic compounds. The curvature of the plots corresponds to the anharmonic terms of the potential energy surface and to the spin-phonon coupling coefficients beyond the first-order. All these features are related to multi-phonons contributions to spin relaxation and their determination is expected to be crucial for the rationalization of spin dynamics in molecular compounds~\cite{Lunghi2019b}. The calculation of first-order spin-phonon coupling coefficients would require more than 1000 CASSCF simulations\cite{Lunghi2017,Lunghi2019b}, a value comparable to the cost of training and testing of the ML model. However, the computational cost of an accurate numerical estimation of second-order coupling coefficients rapidly diverges and exceeds the cost of generating the ML model. This calculation has never been attempted before because of its computational demand and it is only possible within an accelerated framework like the one proposed here.

The method presented here can be generally applied to explore the conformational space of compounds in the search for their optimal properties, either scalar or tensorial. The approach can be readily applied to the prediction of any local atomic quantity of both solid-state materials or isolated molecules as long as an appropriate electronic structure method for the preparation of the training/test sets exists. Regarding magnetism, while solid-state anisotropy is expected to be easily accounted for with the proposed method, an interesting challenge is posed by the modelling of exchange coupling constants because of their non-local nature. We also anticipate that this approach can be extended to the exploration of the entire chemical space, once the scheme is combined with high-throughput electronic structure theory\cite{Curtarolo2013} and generative models \cite{Sanchez-lengeling2018}, such as Variational Autoencoder\cite{Gomez-Bombarelli2018} and Reinforcement Learning\cite{Popova2018}. These are currently limited to the generation of new organic molecules defined by their covalent bonds’ topology. The extension of such generative schemes to the generation of 3-dimensional structures and their combintion with the present approach will represent an important step forward towards the creation of a unified machine learning approach for the exploration of the conformational, configurational and compositional chemical space.

\vspace{0.2cm}
The authors declare no competing interests. This work has been sponsored by Science Foundation Ireland (grant 14/IA/2624). Computational resources were provided 
by the Trinity Centre for High Performance Computing (TCHPC) and the Irish Centre for High-End Computing (ICHEC). We also acknowledge the MOLSPIN COST action CA15128.

\clearpage

\section*{Supplementary Information}

\subsection*{Anisotropy Decomposition in Spherical Harmonics}

The decomposition of the Cartesian tensor $D_{ij}$ into a 2-rank spherical harmonics $T^{m}$, with $m=-2,2$, is done accordingly to the relations
\begin{equation}
 T^{0}=\frac{1}{\sqrt{6}}(3D_{33}-(D_{11}+D_{22}+D_{33}))\:,
\end{equation}
\begin{equation}
 T^{-1}=\frac{1}{2}(D_{13}+D_{31}-i(D_{23}+D_{32}))\:,
\end{equation}
\begin{equation}
 T^{1}=-\frac{1}{2}(D_{13}+D_{31}+i(D_{23}+D_{32}))\:,
\end{equation}
\begin{equation}
 T^{-2}=\frac{1}{2}(D_{11}-D_{22}-i(D_{12}+D_{21}))\:,
\end{equation}
\begin{equation}
 T^{2}=\frac{1}{2}(D_{11}-D_{22}+i(D_{12}+D_{21}))\:.
\end{equation}

\subsection*{Ab Initio Calculations}

\noindent
The ORCA software~\cite{Neese2012} has been employed for all the calculations. We have used the basis sets def2-TZVP 
for C, N and S species and the def2-SVP for C and H species. The def2-TZVP/C auxiliary basis set has been used for all the elements. The calculations of the
$\mathbf{D}$ tensor have been carried out at the CASSCF level of theory, with a (7,5) active space and spin-orbit contributions included through quasi-degenerate perturbation theory. The calculations of the conformational energy has been performed at the DFT level with the PBE functional~\cite{Perdew1996}.

\subsection*{Supervised Learning.} 

\noindent
The coefficients $\alpha$ of the machine learning model where determined by linear Ridge regression: 
\begin{equation}
\min_{ \{ \alpha_{j} \}}
\left[ \| T^{m}_\mathrm{QM}(\{r_{i}\}) -  T^{m}_\mathrm{ML}(\{r_{i}\},\{\alpha^{m}_{j}\}) \|^{2} + \lambda \| \{ \alpha^{m}_{j} \} \|^{2} \right]\:.
 \label{ridge}
\end{equation}
where the first term corresponds to the canonical least-square-fitting of the $T^{m}_{\mathrm{QM}}$ first principles reference values, and the second one to the regularization term. The optimal value of $\lambda$ was chosen as to minimise the error on the validation set. The code LAMMPS~\cite{Plimpton1995} has been used to generate the bi-spectrum components. In all cases the order $2J=8$ for the bi-spectrum components, corresponding to 56 elements per atomic species, has been used. The number of atomic species is a variable that can be adapted to increase the accuracy of the model and does not necessarily need to correspond to the chemical elements. The regression of Co(pdms)$_{2}$'s energy was computed by increased the number of atomic species to nine by discriminating chemically-inequivalent chemical elements. Conversely, the regression of Co(pdms)$_{2}$'s magnetic anisotropy was carried out by only considering the atoms within the radial cutoff distance from the Co atom. In the latter case the correspondence between atomic species and chemical elements was used. The radial cutoff $R_{cut}$ used to build the bi-spectrum components have been optimised as to minimise the overall error on the training/validation set and fixed to 3.5 \AA$ $ for Co(pdms)$_{2}$'s magnetic anisotropy, 3.1 \AA$ $ for Co(pdms)$_{2}$'s energy and 4.5 \AA$ $ for both FeCl$_{x}$'s energy and magnetic anisotropy. The definition of bi-spectrum components gives the possibility to differentiate atomic kinds with weights and atomic radii\cite{Thompson2014}. In this work we have set all the weights to unity and kept all the atomic radii equal to 0.5. The latter condition corresponds to using the same $R_{cut}$ for every species. The covariancy of Ridge regression for tensorial properties requires the estimation of the amount of rotation between each configuration and the reference molecular orientations. This was estimated by applying the Eckart-Sayvets conditions\cite{Neto2006}. This approach provide the rotation matrix that brings the Cartesian coordinates of a structure into those of a rotated one by taking into account that a rigid translation and an intra-molecular motion might also have occurred. This rotation matrix can be interpreted in terms of Euler's angles. The latter are then used to compute the Wigner rotation matrix that appears in Eq. 5 of the main text.

\subsection*{Particle Swarm Optimization.} 

\noindent
The $\alpha-th$ particle in the swarm corresponds to a vector $p^{\alpha}$ that stores the position of every atom in space. The vectors $p^{\alpha}$ are propagated by summing them with the particle velocity $v_{\alpha}$. The velocity of each particle in the swarm was updated at each $i-th$ step with a simple scheme: $v_{\alpha}^{i+1}=\omega v_{\alpha}^{i} + \Gamma[c_{1}p^{\alpha}_{best}+c_{2}p_{best}]$. $p^{\alpha}_{best}$ corresponds to the vector $p^{\alpha}$ that scored the best in the history of the particle $\alpha$, while $p_{best}$ correspond to the vector that scored the best among all the particle in the swarm. The coefficients $\omega$ and $\Gamma$ were chosen as 0.7 and 1.70, respectively. The coefficients $c_{1}$ and $c_{2}$ are random number in the range [0:1]. Tests with different values of $\gamma$ and number of particles have been carried out with no significant difference in the results.

\end{document}